\documentclass[final,3p,times,twocolumn]{elsarticle}
\usepackage{amssymb}
\biboptions{sort&compress}
\newcounter{bla}

\usepackage{graphicx}
\usepackage{bm}
\usepackage{bbm}
\usepackage{slashed}
\usepackage{amsmath}
\usepackage{enumitem}
\usepackage{hhline}
\usepackage{multirow}
\usepackage[utf8]{inputenc} 
\usepackage[dvipsnames]{xcolor}
\usepackage[colorlinks,citecolor=blue]{hyperref}
\usepackage{float}
\usepackage{soul}
\usepackage{algorithm}
\usepackage{algorithmic}
\newcommand{\tab}[0]{\hspace{0.5cm}}

    {}%
    
\definecolor{colcorr}{RGB}{240, 128, 128}

\newcommand{\R}{\varrho}
\newcommand{\gf}{G_{\text{F}}}
\newcommand{\cosenu}{\texttt{COSE$\nu$}}
\journal{Computer Physics Communications}

\begin{document}

\begin{frontmatter}
\title{COSE$\nu$: A Collective Oscillation Simulation Engine for Neutrinos}
\author[a]{Manu George\corref{fauthor}}
\author[b]{Chun-Yu Lin}
\author[a,c]{Meng-Ru Wu}
\author[a]{Tony G. Liu}
\author[d]{Zewei Xiong}

\cortext[fauthor] {Corresponding author.\\\textit{E-mail address:} manug@gate.sinica.edu.tw}

\address[a]{Institute of Physics, Academia Sinica, Taipei, 11529, Taiwan}
\address[b]{National Center for High-performance computing, National Applied Research Laboratories, Hsinchu Science Park, Hsinchu City
30076, Taiwan}
\address[c]{Institute of Astronomy and Astrophysics, Academia Sinica, Taipei, 10617, Taiwan}
\address[d]{GSI Helmholtzzentrum f\"ur Schwerionenforschung, Planckstra{\ss}e 1, 64291 Darmstadt, Germany}

\begin{abstract}
We introduce the implementation details of the simulation code \cosenu, which numerically solves a set of non-linear partial differential equations that govern the dynamics of neutrino collective flavor conversions. 
We systematically provide the details of both finite difference method supported by Kreiss-Oliger dissipation and finite volume method with seventh order weighted essentially non-oscillatory scheme. 
To ensure the reliability of the code, we perform comparison of the simulation results with theoretically obtainable solutions.
In order to understand and characterize the error accumulation behavior of the implementations when neutrino self-interactions are switched on, we also analyze the evolution of the deviation of the conserved quantities for different values of simulation parameters. 
We report the performance of our code with both CPUs and GPUs. 
The public version of the \cosenu~package is available at \url{https://github.com/COSEnu/COSEnu}.
\end{abstract}

\end{frontmatter}

\section{Introduction}
Neutrinos are the most elusive fundamental particles in the Standard Model of particle physics.
They have no electric charge and tiny masses (not yet known experimentally), and only interact with other particles via the weak force. 
Despite the weakly-interacting nature, 
they play a substantial role in deciding the evolution and the matter composition of different physical systems such as  core-collapse supernovae (CCSNe), neutron star (NS) mergers, and the early Universe. Added to their elusiveness, terrestrial based experiments and astrophysical observations revealed that neutrinos are capable of undergoing a phenomenon known as flavor transition, by virtue of which they oscillate between $e$, $\mu$, and $\tau$ flavors, due to the mixing of their flavor and mass quantum eigenstates~\cite{Zyla:2020}. 
Associated with these properties comes a set of parameters: 
the masses of different mass eigenstates 
and their ordering, as well as the mixing angles and the CP-violating phase(s) in the neutrino mixing matrix. 
Most of these parameters have been measured precisely in recent years~\cite{Zyla:2020} while the ongoing and planned experiments are expected to shed more light on the remaining ones.

Recent theoretical studies revealed that neutrinos behave differently when surrounded by dense matter. 
Apart from the Mikheyev–Smirnov–Wolfenstein (MSW) \cite{Wolfenstein:1979, MikSmi:1985} mechanism in which neutrinos experience resonant flavor transition due to their neutral and charged current interactions with matter, they can also undergo forward scattering among themselves via neutral current interactions resulting in processes such as $\nu_e \bar\nu_e \rightleftharpoons \nu_x \bar\nu_x$~\cite{Pantaleone:1992eq,Sigl:1992fn}, where $\nu_x$($\bar\nu_x$) represent $\nu_\mu$($\bar\nu_\mu$) or $\nu_\tau$($\bar\nu_\tau$) neutrinos, leading to the collective conversion of neutrino flavors; see Ref.~\cite{Capozzi:2022slf} for a recent review and references therein.
The conversion rates associated with this mechanism can be as fast as 
$\omega_{\text{coll}}\propto \gf n_{\nu_e}$, where $\gf$ is the Fermi constant and $n_{\nu_e}$ is the local number density of $\nu_e$. 
For typical neutrino densities ($10^{30}-10^{35}\text{cm}^{-3}$) around the neutrino emission surfaces of CCSNe and coalescing NSs, 
the corresponding $\omega_{\text{coll}}\sim \mathcal{O}(\text{cm}^{-1})$ greatly exceeds the vacuum oscillation rate $\omega_{\text{vac}}= |\Delta m^2|/2E \sim \text{km}^{-1}$, where $\Delta m^2$ is the mass-squared difference of the relevant neutrino mass eigenstates and $E$ is the neutrino energy. 
Various analytical and numerical studies have been carried out to  understand the behavior of these ``fast'' neutrino flavor conversions~\cite{Sawyer:2016,Izaguirre:2017,Dasgupta:2017,Capozzi:2017,Abbar:2018,Capozzi:2019,Yi:2019,Chakraborty:2019wxe,Joshua:2020,Bhattacharyya:2020,Johns:2020,Padilla-Gay:2020uxa,Capozzi:2020kge,Bhattacharyya:2021,Richers:2021nbx,Wu-COSEnu-1:2021,Morinaga:2021,Zewei:2021,Zaizen:2021wwl,Kato:2021cjf,Sherwood:2021,Dasgupta:2021gfs,Abbar:2021lmm}. 
Besides, several studies on systems such as the NS merger and CCSNe~\cite{Wu:2017-05,Wu:2017-12,Morinaga:2019wsv,Nagakura:2019sig,Johns:2020,DelfanAzari:2019tez,Abbar:2019zoq,Glas:2019ijo,Zewei:2020,Manu:2020,Li:2021,Nagakura:2021hyb} have inferred that the fast conversions may occur in the most dense regions of them and can potentially affect the dynamics and the nucleosynthesis of these systems. 

Because of the highly non-linear structure of the neutrino-neutrino effective interaction Hamiltonian, it is impossible to have a complete analytical solution for the neutrino flavor evolution in environments dense in neutrinos. 
However, it is possible to linearize the system of equations and carry out the normal mode analysis~\cite{Banerjee:2011fj,Izaguirre:2017,Capozzi:2017,Yi:2019}. 
Such analyses show that dense neutrino system supports collective flavor runaway modes when certain criteria are satisfied. 
Although the linear analysis helps us understand the behavior of fast oscillations to some extent, the assumption that the correlations between different flavors are much smaller compared to the self correlation put a very stringent constraint on the applicability of this method. 
Thus, to study the neutrino oscillations in dense environments to full extent, several numerical simulations have been carried out in recent years.  
Refs.~\cite{Joshua:2020, Bhattacharyya:2021, Wu-COSEnu-1:2021} studied the neutrino system in a one-dimensional (1D) box in the $z$-direction with periodic boundaries and translational symmetry in the $x$ and $y$ directions using numerical methods which evolves directly neutrino correlations in discretized grids. 
Ref.~\cite{Zaizen:2021wwl} used the spectral method to solve the same set of equation in 1D.
Meanwhile, the authors of Ref.~\cite{Sherwood:2021} adopted the particle-in-cell method and have recently performed simulations for systems with higher spatial dimensions, including both 2D and 3D cases.
A study which compares in detail simulation outcome from these groups will be published soon~\cite{codecomparison}. 

The intention of this article is to systematically discuss the computational aspects of \cosenu~(Collective Oscillation Simulation Engine for Neutrinos), the code base developed for simulating the collective 
neutrino flavor conversions and used in Ref.~\cite{Wu-COSEnu-1:2021}. 
\cosenu~is written completely in \texttt{C++} to ensure the high performance. 
At present, it provides two advanced numerical methods for solving the neutrino flavor evolution equations. 
The first one uses the 4th order finite-difference (central) scheme (FD) supported by third-order Kreiss-Oliger (KO3) numerical dissipation to treat the advection effectively. 
The second method adopts the finite-volume (FV) formalism. 
In this scheme, the flux reconstruction is implemented using the 7th order weighted essentially non-oscillatory (WENO) scheme. 
For both cases, we have used the 4th order Runge-Kutta method (RK4) for time integration. 
We shall provide implementation details of both methods in the following sections. 
In-depth analysis of the physics of the simulation results and their comparison with previous numerical studies for some bench-mark cases was reported in Ref.~\cite{Wu-COSEnu-1:2021}.

This article is organized as follows. 
In section~\ref{sec:theo_setup}, we briefly discuss the equations governing the neutrino flavor evolution.
Section~\ref{sec:numerics} is dedicated to detailed descriptions of the numerical schemes used in \cosenu. 
In section~\ref{sec:tests}, we summarize the results from the advection tests which provide some insights into the numerical error accumulation behavior of both FD and FV implementations. 
We also carry out tests when the vacuum oscillation is present and compare the results with the analytical solutions. 
In section~\ref{sec:coll_osc_test}, we present the results from the simulations when collective neutrino flavor transitions occur. 
We provide the computational performance comparison in section~\ref{sec:performance}. 
Discussion and conclusions follow in section~\ref{sec:discussion}. 
Throughout this paper we set $\hbar=c=1$ to adopt the natural units.


\section{Theoretical setup}
\label{sec:theo_setup}

We considered a simplified set of quantum kinetic equations to describe the transport of neutrinos in a physical system~\cite{Vlasenko:2014,Volpe:2013}. 
First, we assume for simplicity a two flavor neutrino system in which the single particle flavor state can be represented by a $2\times 1$ column vector $(\nu_e, \nu_x)^T$, where $\nu_e$ represents the electron type neutrino and $\nu_x$ represent the $\mu$ type or $\tau$ type neutrino or an appropriate linear combination of them. 
Then, we use a $2\times 2$ complex valued (Hermitian) density matrix $\R(t,\bm x, \bm v)$ [$\bar\R(t,\bm x, \bm v)$] to represent 
the quantum statistical phase-space distribution of (anti-)neutrinos in our system. 
The diagonal elements of $\R$, labelled as $\R_{ee}(t,\bm x, \bm v)$ and $\R_{xx}(t,\bm x, \bm v)$ are related to the number densities of $e$ and $x$ type neutrinos respectively for a given velocity mode $\bm v$. 
The off-diagonal component $\R_{ex}(t,\bm x, \bm v)$ is related to the correlation between them. 
The same definition applies to the antineutrino density matrix $\bar\R$.
For this first version of \cosenu, we choose to ignore other potential correlations such as spin coherence and the neutrino-antineutrino correlators~\cite{Vlasenko:2014,Volpe:2013}, as well as the momentum-changing collisions of neutrinos~\cite{Shalgar:2020wcx,Johns:2021,Martin:2021xyl,Sigl:2021tmj}. 
We also assume that the system has a perfect translational symmetry in both $x$ and $y$ directions and axial symmetry about the $z$-axis such that  
the density matrices depend only on $t$, $z$ and the $z$ component of the velocity $v_z$. 
Then, the space-time evolution of $\R(t, z, v_z)$ and $\bar\R(t, z, v_z)$ are determined by the following equations in 1+1+1(time+space+velocity) dimensions.
\begin{subequations}\label{eq:eom}
\begin{align}
& \frac{\partial}{\partial t}\R(t,z,v_z)+
v_z\frac{\partial}{\partial z}\R(t,z,v_z)
=-i[H(t,z,v_z),\R(t,z,v_z)],\label{eq:eom_nu}\\
& \frac{\partial}{\partial t}\bar\R(t,z,v_z)+
v_z\frac{\partial}{\partial z}\bar\R(t,z,v_z)
=-i[\bar H(t,z,v_z),\bar\R(t,z,v_z)], \label{eq:eom_anu}
\end{align}
\end{subequations}

\noindent
where the square bracket ($[,]$) stands for the commutation operation. 
The explicit form for $\R$ (for neutrinos) and $\bar\R$ (anti-neutrinos) are given by, 
\begin{equation}\label{eq:rho}
\varrho(t,z,v_z)=
\left [
\begin{array}{cc}
\R_{ee} & \R_{ex}\\
\R_{ex}^* & \R_{xx} \\
\end{array}
\right ],~
\bar\R(t,z,v_z)=
\left [
\begin{array}{cc}
\bar\R_{ee} & \bar\R_{ex}\\
\bar\R_{ex}^* & \bar\R_{xx} \\
\end{array}
\right ].
\end{equation}

The Hamiltonian $H$ (and $\bar H$) in general have contributions stemming from the vacuum oscillations ($H_{\text{vac}}$), neutrino-matter forward scattering ($H_{\text{m}}$) and neutrino-neutrino forward scattering ($H_{\nu\nu}$). 
In the numerical setup we omit the neutrino-matter forward scattering contribution as it can be removed, assuming $H_{\text{m}}$ is space-time independent over the scales determined by $H_{\nu\nu}$, by making appropriate transformations of the density matrices~\cite{Duan:2006-12,Duan:2010bg}\footnote{This amounts to the redefinition of the frames in the flavor space.
Note that under such a transformation, the vacuum Hamiltonian becomes an effective one ($H_{\rm vac}\rightarrow H_{\rm vac}^{\rm eff}$). For simplicity, we omit the superscript ``eff'' hereafter.}. 
In that case the Hamiltonian relevant for our present discussion takes following form in the flavor basis, 

\begin{subequations}\label{eq:ham}
\begin{equation}\label{eq:ham_nu}
\begin{split}
 H(t,z,v_z)= H_{\rm{vac}}
+ &\mu \int_{-1}^1 dv_z' (1-v_z v_z')\\&
\times[\R(t,z,v_z')-\alpha\bar\R^*(t,z,v_z')],
\end{split}
\end{equation}
 \begin{equation}\label{eq:ham_anu}
 \begin{split}
 \bar H(t,z,v_z)= \bar H_{\rm{vac}}
-&\mu \int_{-1}^1 dv_z' (1-v_z v_z')\\&
\times [\R^*(t,z,v_z')-\alpha \bar\R(t,z,v_z')],
\end{split}
\end{equation}
\end{subequations}
\noindent
where 
\begin{equation*}
    H_{\text{vac}}=\bar H_{\text{vac}}=\frac{\omega_{\rm vac}}{2}\left [\begin{array}{cc}
    -\cos 2\theta_{\rm{vac}} &\sin 2\theta_{\rm{vac}}\\
    \sin 2\theta_{\rm{vac}} &\cos 2\theta_{\rm{vac}}
    \end{array}\right ]
\end{equation*}

\noindent
is the vacuum mixing Hamiltonian with $\theta_{\rm vac}$
the mixing angle. 
The quantity $\mu = \sqrt{2}\gf n_{\nu_e}^0$ represents the effective strength of $H_{\nu\nu}$, where $n_{\nu_\beta}^0$ denotes the number density of $\beta$ flavor neutrinos.   $\alpha=(n_{\nu_e}^0/n_{\bar\nu_e}^0)$ is the initial  neutrino-antineutrino asymmetry parameter.
Note here that the quantities $\R$ and $\bar\R$ are normalized with respect to the initial number densities  
of $\nu_e$ and $\bar\nu_e$ respectively.

In literature, another equivalent way of describing the dynamics of the collective neutrino flavor evolution is via the so-called polarization vectors, $\bm P$, defined by
\begin{equation}
 \rho = (P_0\mathcal{I} + \bm P\cdot\bm\sigma)/2,
\end{equation}

\noindent
where $\bm\sigma$ are the Pauli matrices.
For cases where the number densities of neutrinos and antineutrinos are conserved per phase-space volume, 
both $P_0$ and the length of the polarization vector $|\bm P| = \sqrt{\sum_{i=1}^3 P_i^2}$ are constants of motion. 
Due to its ease of implementation and also keeping potential future extensions in mind, \cosenu~is implemented in terms of the density matrix formalism.
We make use of the conserved quantities while testing simulation (see Sec.~\ref{sec:coll_osc_test}).


\section{Numerical implementation}
\label{sec:numerics}

In this section we focus on the numerical implementation of \cosenu~which solves the partial differential equations (PDEs) (\ref{eq:eom_nu}) and (\ref{eq:eom_anu}) when supplemented with the Hamiltonian in Eq.~(\ref{eq:ham}). 
The idea in both FD and FV implementations is to divide the spatial domain which extends from $z_0$ to $z_1$ into $N_z$ grid points and carry out the time integration using RK4 from an initial time $t_0$ to a final time $t_f$ in steps of $\Delta t$. 
Then the size of each cell $\Delta z = (z_1-z_0)/N_z$ is related to $\Delta t$ through $\Delta t=C_{\text{CFL}}(\Delta z/\text{max}(|v_z|))$, where $C_{\text{CFL}}$ is the Courant–Friedrichs–Lewy number.
Since the interaction Hamiltonian is velocity dependent and $v_z$ takes values from $-1$ to $1$, the velocity space is also divided into $N_{v_z}$ discrete points such that the neutrino beams with $v_z$ values ranging between $v_j$ and $v_j + \Delta v_z$ for some $0\leq j < N_{v_z}$, are treated as a single beam with $v_z$ value $v_j+\Delta v_z/2$. Then the contribution to the Hamiltonian from the velocity integrals in the Eqs.~\eqref{eq:ham_nu} and \eqref{eq:ham_anu} are carried out using simple Riemann sum.


\subsection{Finite difference method}
\label{sec:fd}
The numerical technique used to solve hyperbolic PDEs in this work is based on the ``method of lines'' (See Sec.~6.7 of Ref. \cite{Gustafsson:1995} for instance) where the spatial and temporal discretizations are treated conceptually in a separated style. 
We discretized the advection term with the 4th order central finite difference scheme. 
The resulting equations become ordinary differential equations (ODEs) of time and can be solved as the initial-value problem via the explicit, 4th order Runge-Kutta method. 
For the integration over velocities in the right hand side, the standard trapezoid rule of second-order accuracy are used with the fixed-width, vertex-center grid points over velocities in $[-1,1]$. 
We have also used the basic Simpson's rule of fourth-order accuracy and obtained similar results. 
The framework of ``method of lines'' allows us to treat the advection term with other high order methods such as the WENO scheme, which will be discussed in the next section.

To suppress the high-frequency instability arising from the FD approximation of equations (see Sec.~\ref{sec:coll_osc_test}), we add Kriess-Oliger dissipation on the right hand side of each evolving variables $u$, i.e., letting $\partial_t u \rightarrow \partial_t u + Qu$.
The general form of the $2r$-order Kreiss-Oliger dissipation can be expressed as

\begin{equation}
Q=\varepsilon_\mathrm{ko} (-1)^r h^{2r-1}(D_+)^r (D_-)^r / 2^{2r},
\end{equation}

\noindent
where $D_\pm$ are one-sided FD operators for $\partial/\partial x$, defined by $D_{+}\equiv (u_{+1}-u_0)/h$ and $D_{-}\equiv (u_{0}-u_{-1})/h $. With this convention, the central FD of $\partial^2/\partial x^2$ can be written as $D_+D_- = (u_{+1}-2u_0+u_{-1})/h^2$. 
A sufficiently large dissipation strength $\varepsilon_\mathrm{ko}$ can suppress the instability without destroying the convergence of a symmetric hyperbolic system of equations, as discussed in \cite{Gustafsson:1995} via the Fourier-based analysis. 
In our work, we have used $r=2$ for most numerical experiments. 


\subsection{Finite volume method}
\label{sec:fv}
Unlike the FD method where the values on the grids points are evolved, the FV scheme evolves the cell-averaged values. Let us consider the following simple one dimensional hyperbolic equation, 
\begin{equation}
    \label{eq:hyper}
    \frac{\partial u(t, z)}{\partial t} + \frac{\partial f(u(t, z))}{\partial z} = 0,
\end{equation}

\noindent
where $u(t, z)$ represents the quantity that we want to evolve, and $f(u(t, z))$ is the associated flux function. 
Given the values of the functions $u$ and $f$ at the grid points and assuming we can appropriately interpolate them to a required order of accuracy, we have
\begin{equation}
    \frac{d \bar u_i(t)}{d t} = -\frac{1}{\Delta z}\left[f_{i+1/2}(t) - f_{i-1/2}(t)\right].
    \label{eq:cell_av}
\end{equation}

In the above expression we have used the definition $\bar A_i(t) = (1/\Delta z)\int_{i-1/2}^{i+1/2} dz A(t, z)$ for some function $A(t, z)$. 
The upper and lower limits of the integral denoted by  $i\pm 1/2$ are the representatives of $z_i\pm(\Delta z/2)$, with $z_i$ being the coordinate of the $i$th grid point. 
The PDE in Eq.~(\ref{eq:hyper}) now becomes an ODE  
in $t$, which can be solved, given the values on the right hand side of Eq.~(\ref{eq:cell_av}), using any standard ODE solver.
Then the problem is down to reconstructing the flux values at the cell boundaries $i\pm 1/2$ to a required order of accuracy. 
For reconstructing the flux values $f_{i\pm 1/2}$, the FV implementation of \cosenu~uses WENO scheme \cite{Liu:1994, Shu:1998, Shu:2003} such that the fluxes are 7-th order accurate for a smooth function while they are at least 4-th order accurate for functions with discontinuities.
The discussion below closely follows Refs. \cite{Liu:1994, Shu:1998}.

Given the cell averaged values of the flux function $f(z)$ at $n$ contiguous locations with a uniform separation of $\Delta z$, we can construct a polynomial $\hat f^{(n)}$ of degree $n-1$ such that the reconstruction of the function $f(z)$ is $n$-th order accurate. That is,
\begin{equation}
    f(z_i) = \hat f^{(n)}_i + O(\Delta z^n).
\end{equation}

This allows us to write down the $n$-th order approximation of Eq.~(\ref{eq:cell_av}) as,
\begin{equation}
    \frac{d}{dt}\bar u_i(t) = -\frac{1}{\Delta z}\left[\hat f_{i+1/2} - \hat f_{i-1/2}\right] + O(\Delta z^n).
    \label{eq:hyp_O_n_desc}
\end{equation}

We have omitted the superscript $n$ in Eq.~\eqref{eq:hyp_O_n_desc}. 
It can be shown that, given the cell averaged values of the function $\bar f_i$ at each cell of the stencil, values of the function at $i\pm 1/2$  can be expressed as a linear combination of the values of $\bar f_i$,
\begin{subequations}
    \begin{equation}
        \hat f_{i+1/2} = \sum_{j=0}^{n-1} c_{r,j} \bar f_{i-r+j},
    \end{equation}
    \begin{equation}
        \hat f_{i-1/2} = \sum_{j=0}^{n-1} \tilde c_{r,j} \bar f_{i-r+j},
    \end{equation}
    \label{eq:flux_inter}
\end{subequations}

\noindent
with the following definition of the coefficients
$c_{rj}$ (see Ref. \cite{Shu:1998} for instance), 
\begin{equation}
    c_{r,j} = \sum_{m=j+1}^k \frac{\sum^k_{l=0, l\neq m} \Pi^k_{q=0, q\neq m,l}(r-q+1)}{\Pi^k_{l=0, l\neq m}(m-l)},
    \label{eq:crj}
\end{equation}

\noindent
where $r$ is the shift parameter that can be thought as the measure of the left-shift of the stencil under consideration from the point $i$. 
Since we are reconstructing the values of $f$ at the exact locations $i\pm 1/2$, the relation $\tilde c_{r,j} = c_{r-1, j}$ holds true. 
Note here that the definition of $c_{r,j}$ in the Eq.~(\ref{eq:crj}) is valid only for a grid with uniform cell size $\Delta z$.

The $n$-th order WENO scheme considers a stencil $S$ containing $n$ grid points in order to approximate the value of flux at $i\pm 1/2$. 
Then this stencil $S$ is divided into smaller sub-stencils of order $k$ such that $n = 2k-1$. 
The purpose of this decomposition is that, the reconstructed values at the given location is $n$-th order accurate for smooth functions and at least $k$-th order accurate  
if there exists a discontinuity. 
Each sub-stencil of length $k$ can then be labelled using the shift parameter $r$ (see Fig. \ref{fig:stencils}). 
As a result, we have
\begin{equation}
    \hat f_{i\pm 1/2}^{(n)} = \sum_{r}d_r \hat f^{(k)}_{r,i\pm 1/2}.
    \label{eq:fn_lin_comb}
\end{equation}
In order to treat the discontinuities appropriately while making the above linear combination [Eq.~\eqref{eq:fn_lin_comb}], WENO uses the weighted averaging of the coefficients $d_r$. 
The weight factor for the contribution of the sub-stencil $r$ is estimated with respect to a smoothness index (SI$_r$), which as the name implies, measures the smoothness of the sub-stencil $r$. In our  implementation, we have used the results from Refs.~\cite{Jiang:1996} and \cite{Balsara:2000} to compute the $\text{SI}_r$. Given the values of $\text{SI}_r$, we replace $d_r$ in Eq.~(\ref{eq:fn_lin_comb}) with the corresponding weighted coefficients,
  
  \begin{equation}
    \hat f_{i\pm 1/2}^{(n)} = \sum_r \frac{w_r}{w}\hat f^{(k)}_{r, i\pm 1/2},
    \label{eq:fn_w_lin_comb}
\end{equation}

\noindent
where $w=\sum_{r} w_r$ and 
\begin{equation}
    w_r = \frac{d_r}{(\text{SI}_r+\varepsilon)^2}.
    \label{eq:weight}
\end{equation}
The $\varepsilon$ in Eq. (\ref{eq:weight}) is to avoid the possibility of the denominator becoming zero. 
A typical value of $\varepsilon$ can be $\sim 10^{-6}$. 
The explicit forms of the quantities $\hat f^{(n)}$, $\hat f^{(k)}$, $d_r$, and $\text{SI}_r$ for 7th order WENO are provided in the~\ref{app:weno7_explicit}. Note that, when S$_r$ has discontinuity $\text{SI}_r$ becomes large resulting in smaller value of $w_r$ which as we can see from the Eq.~\eqref{eq:fn_w_lin_comb} reduces the contribution of $\hat f^{(k)}_{r, i\pm 1/2}$ to $\hat f_{i\pm 1/2}^{(n)}$.


\section{Tests for advection and vacuum oscillations}
\label{sec:tests}

Before discussing collective neutrino oscillations 
in the next section, we use this section to discuss the results of some preliminary tests carried out with \cosenu~when $H_{\nu\nu}$ is turned off. 
In Sec.~\ref{subsec:Advection} we illustrate the results only with pure advection and in Sec.~\ref{subsec:vac_osc} we discuss the results when vacuum oscillation term is turned on. 
In both cases we will show that the simulation results match very well with what expected theoretically.  
Also, to quantify the numerical error, we define the quantity 

\begin{equation}
    E_2 = \sqrt{\frac{1}{L} \int dz \epsilon(z)^2, }\label{eq:E2}
\end{equation}

\noindent
where $L$ is the length of spatial domain and $\epsilon(z) =  f_\text{Exact}(z)-f_\text{Sim}(z)$ with subscripts $_{\rm Exact}$ and $_{\rm Sim}$ denoting results from simulations and from the analytical solutions. 
In the following discussions we use FD and FV as the abbreviations for the fourth order finite-difference method with third order KO dissipation and finite volume method with seventh order WENO flux reconstruction schemes, respectively. 
Note that we adopt the periodic boundary condition in $z$ for all the simulations discussed later. 

\subsection{Advection}
\label{subsec:Advection}

To test the the numerical implementation of the advection term, we set the right hand side of the Eq.~(\ref{eq:eom}) to zero. 
The resulting PDE has the solution of the form $f(z-v(t-t_0))$ for an initial profile $f(z, t_0)$ when the advection velocity is set to $v$. 
To illustrate the behavior of the advection implementations, we consider two different initial profiles $f(z, t_0)$:
a unit Gaussian to represent continuous profiles and a boxcar profile of unit height representing profiles with discontinuities. 

\begin{figure*}[ht]
    \includegraphics[width=1\textwidth]{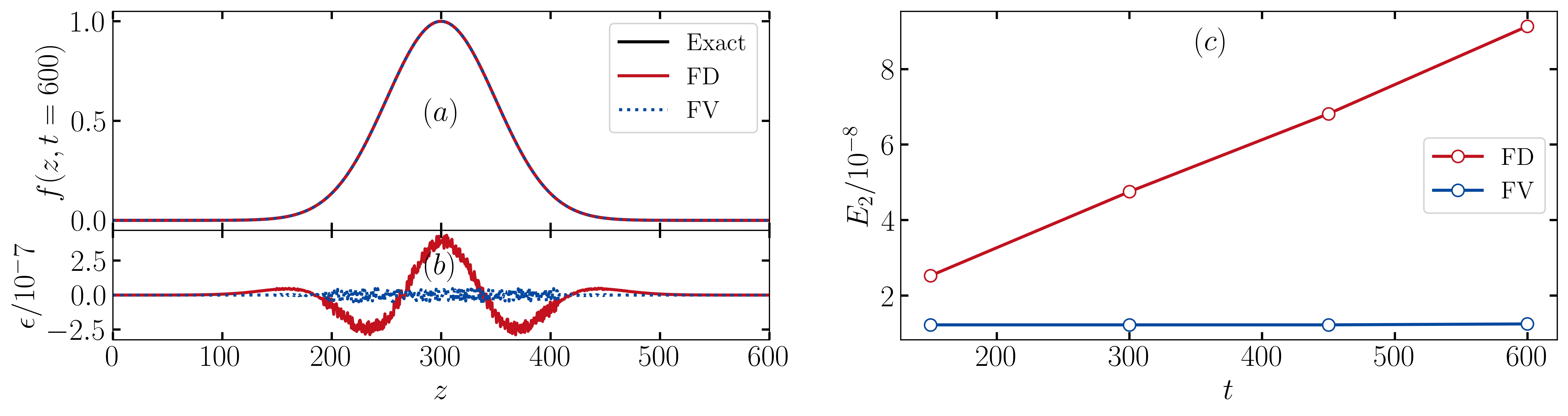}
    \caption{
    Results of the advection tests 
    for a traveling Gaussion packet initially centered at $z=0$ with velocity $v_z=0.5$.
    Panel (a): numerical solutions obtained with FV and FD schemes as well as the exact solution at $t=600$. 
    Panel (b): the spatial distribution of the error $\epsilon(z)$ at the same $t = 600$ as in panel (a).
    Panel (c): the error indicator $E_2$ (see text for definition)  obtained with FV and FD at different times during the evolution.
    }
    \label{fig:adv_test_eps}
\end{figure*}
Figure~\ref{fig:adv_test_eps} shows the results of the  advection tests with the Gaussian profile of unit amplitude initially centered at $z=0$. 
We have chosen $N_z = 2000$, $C_\text{CFL} = 0.2$ and $v=0.5$ mode.
The panel (a) shows the exact solution and the results from both FV and FD at the end of the simulation at $t=600$ while panel (b) shows the corresponding errors. 
Here we see that both implementations of the advection are capable of producing the expected profile with errors at the order of $\mathcal{O}(10^{-8}$). 
The panel (c) demonstrates the error accumulation behavior of FD and FV. It is evident 
that even though the magnitudes of the errors remain 
small, FD accumulates the errors
faster compared to FV. The fact that the error remains almost constant in FV reflects the conservative nature of the FV scheme.

\begin{figure*}[ht]
    \includegraphics[width=\textwidth]{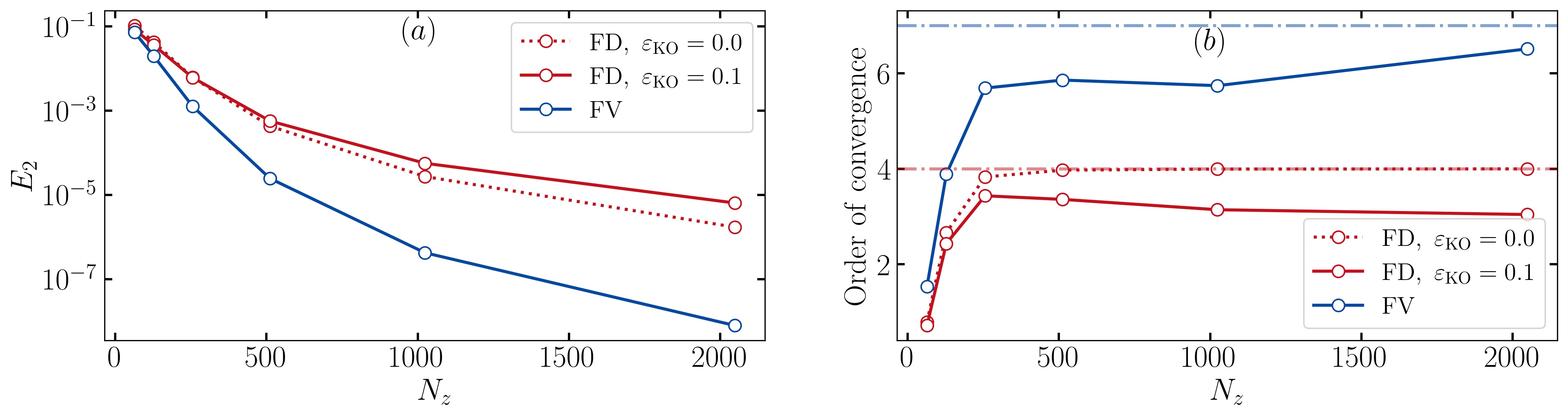}
    \caption{
    $E_2$ [panel (a)] at t=600 and the corresponding
    convergence rate [panel (b)] from advection tests for different values of $N_z$.
    Blue and red lines are used to represent the results obtained with FV and FD schemes, respectively. 
    The dotted and solid red lines distinguishes the results with and without the Kriess-Oliger dissipation. 
    The light dot-dashed blue and red lines in panel (b) represent the theoretical asymptotic values for the order of convergence.  
    }
    \label{fig:adv_test_err}
\end{figure*}

In panels (a) and (b) of Figure~\ref{fig:adv_test_err}, we show the $E_2$ errors and the corresponding numerical order of convergence (see \ref{app:numerr} for more details)  respectively for different values of $N_z$. 
Blue colored lines are used to indicate the results from FV while red colored lines (dotted and solid lines for results with and without Kriess-Oliger dissipation respectively) are used for FD.
As can be observed from Figure~\ref{fig:adv_test_err}, the errors in  FD and FV reduce with increasing value of $N_z$. 
The panel (b) shows that both implementations asymptotically reach the expected order of accuracy (7 for 7th order WENO and 4 for 4th order FD for a smooth function). 
Note that the numerical accuracy gets reduced when we include the KO dissipation for FD (for the tests we have used KO3 with $\varepsilon_\text{KO} = 0.1$). 
However, this sacrifice in the numerical accuracy is compensated by the numerical stability as will be illustrated in the section~\ref{sec:coll_osc_test}. 

Next, we demonstrate the nature of the \cosenu~implementation when some part of the domain contains discontinuity. 
For this purpose, we have chosen an initial box car profile,
\begin{equation}
    f(z, t=0) = \begin{cases}
        1 & \text{for } -100 \le z\le 100,\\
        0 & \text{otherwise.}
    \end{cases}
    \label{eq:box}
\end{equation}

\begin{figure*}[ht]
    \includegraphics[width=\textwidth]{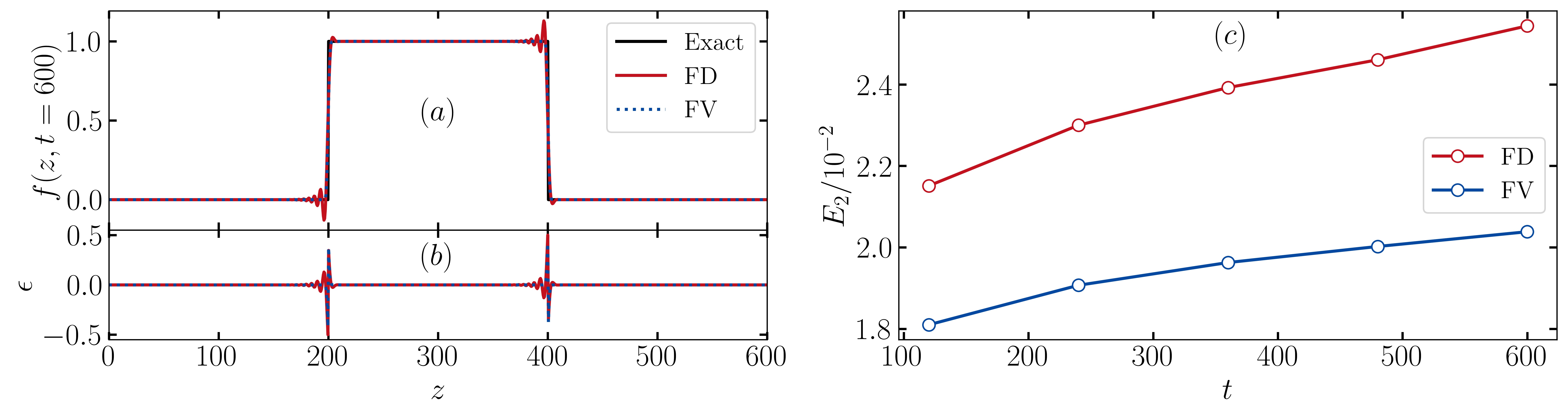}
    \caption{
   Results of advection tests with a box centered initially at $z=0$ with $v_z=0.5$. 
   Panel (a)--(c) show the same quantities as in Figure~\ref{fig:adv_test_eps}.  
   }
    \label{fig:adv_step}
\end{figure*}

The results obtained are shown in Figure~\ref{fig:adv_step}. 
The sub-panel (a) shows the comparison of the results from FD and FV with the exact solution. 
The sub-panel (b) shows the nature of error accumulation of both FD and FV. 
Unlike the Gaussian profile, the maximum values of the error are $\sim 0.5$ throughout the evolution for both cases. 
This is pertained to the slight increase in the width at the base and a slight decrease in the width at the top of the profile obtained from the simulation compared to the exact one. 
Furthermore, the FD scheme produces small oscillation errors across the discontinuities due to the Gibbs phenomenon \cite{Gustafsson:1995}. When increasing the resolution, we find that the maximum value of the errors in both FD and FV across the discontinuities remains unchanged. 
However, the $E_2$ error can be reduced with larger $N_z$. 

\begin{figure*}[ht]
    \includegraphics[width=\textwidth]{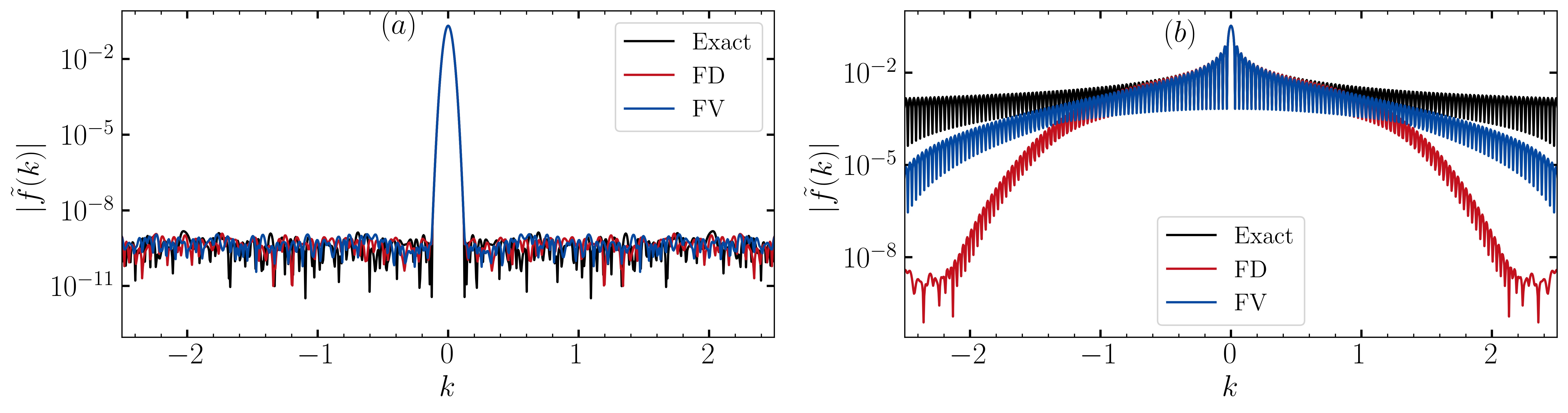}
    \caption{
    Comparison of the discrete Fourier spectrum of the solutions obtained using FD and FV simulations with that of the exact solutions at $t=600$. 
    Panel (a): cases with initial Gaussian profile shown in Figure~\ref{fig:adv_test_eps}. 
    Panel (b): cases with initial box car profile shown in Figure~\ref{fig:adv_step}. 
    }
    \label{fig:fft_adv}
\end{figure*}

We additionally compute the Fourier transformed spectra for all cases discussed above.
In Figure~\ref{fig:fft_adv}, panel (a) compares the discrete Fourier spectrum of the exact solution with the same of the simulation result for an initial Gaussian profile. Similarly panel (b) illustrates those for an initial box profile.
As evident from the figure, the simulations produce nearly identical Fourier spectrum 
as the exact one when the Gaussian profile is used.
For cases with the box profile,  
the Fourier spectra at the low frequencies compares well with the exact one for both the FD and FV schemes. 
Both implementations clearly suppress the  
high frequency part of the Fourier spectrum 
due to the slight deviation from perfect discontinuities at the edges of the box.
Moreover, the FD scheme here shows further suppression of the high $|k|$ power when compared to the FV scheme,   
as a result of the KO dissipation discussed in the section~\ref{sec:fd}. 
A comparison of the advection test results with and without the KO dissipation for FD is shown in~\ref{app:KO}.

\subsection{Vacuum oscillations}
\label{subsec:vac_osc}

The second set of tests that we do for \cosenu~is to consider the cases including both the advection and the vaccum oscillations of neutrinos. 
Since the vacuum Hamiltonian $H_\text{vac}$ is independent of $\R$,  
this allows us to obtain an analytical solution for the space-time evolution of $\R(t, z)$. 
For a pure initial state consisting of only electron neutrinos, the analytical solution of the components of density matrix under the action of $H_\text{vac}$ takes the following form,
\begin{subequations}
\label{eq:vac_sols}
\begin{equation} 
\begin{split}
    \R_{ee}(t, z) = &\R_{ee}^{t_0}(z-v(t-t_0))\\ \times & \left[1-\text{sin}^2(2\theta_\text{vac})\text{sin}^2(\frac{\omega_\text{vac} }{2}(t-t_0))\right],
\end{split}
\end{equation}
\begin{equation} 
\begin{split}
    \R_{xx}(t, z) = & \R_{ee}^{t_0}(z-v(t-t_0))\\ \times &
    \left[\text{sin}^2(2\theta_\text{vac})\text{sin}^2(\frac{\omega_\text{vac} }{2}(t-t_0))\right],
\end{split}
\end{equation} 
\begin{equation}
\begin{split}
    \text{Re}[\R_{ex}(t, z)] =& -\R_{ee}^{t_0}(z-v(t-t_0)) \\ \times &
    \left[\text{sin}(2\theta_\text{vac})\text{cos}(2\theta_\text{vac})  \text{sin}^2(\frac{\omega_\text{vac} }{2}(t-t_0))\right],
\end{split}
\end{equation}
\begin{equation} 
\begin{split}
    \text{Im}[\R_{ex}(t, z)] =&\R_{ee}^{t_0}(z-v(t-t_0))\\
    \times & \frac{1}{2}\left[\text{sin}(2\theta_\text{vac})\text{sin}(\omega_\text{vac} (t-t_0))\right], 
\end{split}
\end{equation}
\end{subequations}

\noindent
where $\R_{ab}^{t_0}(z)$ is the value of the of density matrix components at $t=t_0$ for the matrix component $ab$. 
Note that here we are only interested in testing if the simulation results match with the exact solutions given in the Eqs.~(\ref{eq:vac_sols}). 
For this purpose, we consider a wave packet of pure electron neutrinos with a Gaussian profile initially centered at $z=0$. 
The entire wave packet travels with $v_z=0.5$ in the $z$-direction. We set $\omega_\text{vac}$ to $0.1$ and $\theta_\text{vac}$ to $37$ degrees and choose $N_z = 2000$ and $C_\text{CFL}=0.2$.
 
 \begin{figure*}[h]
    \includegraphics[width=\textwidth]{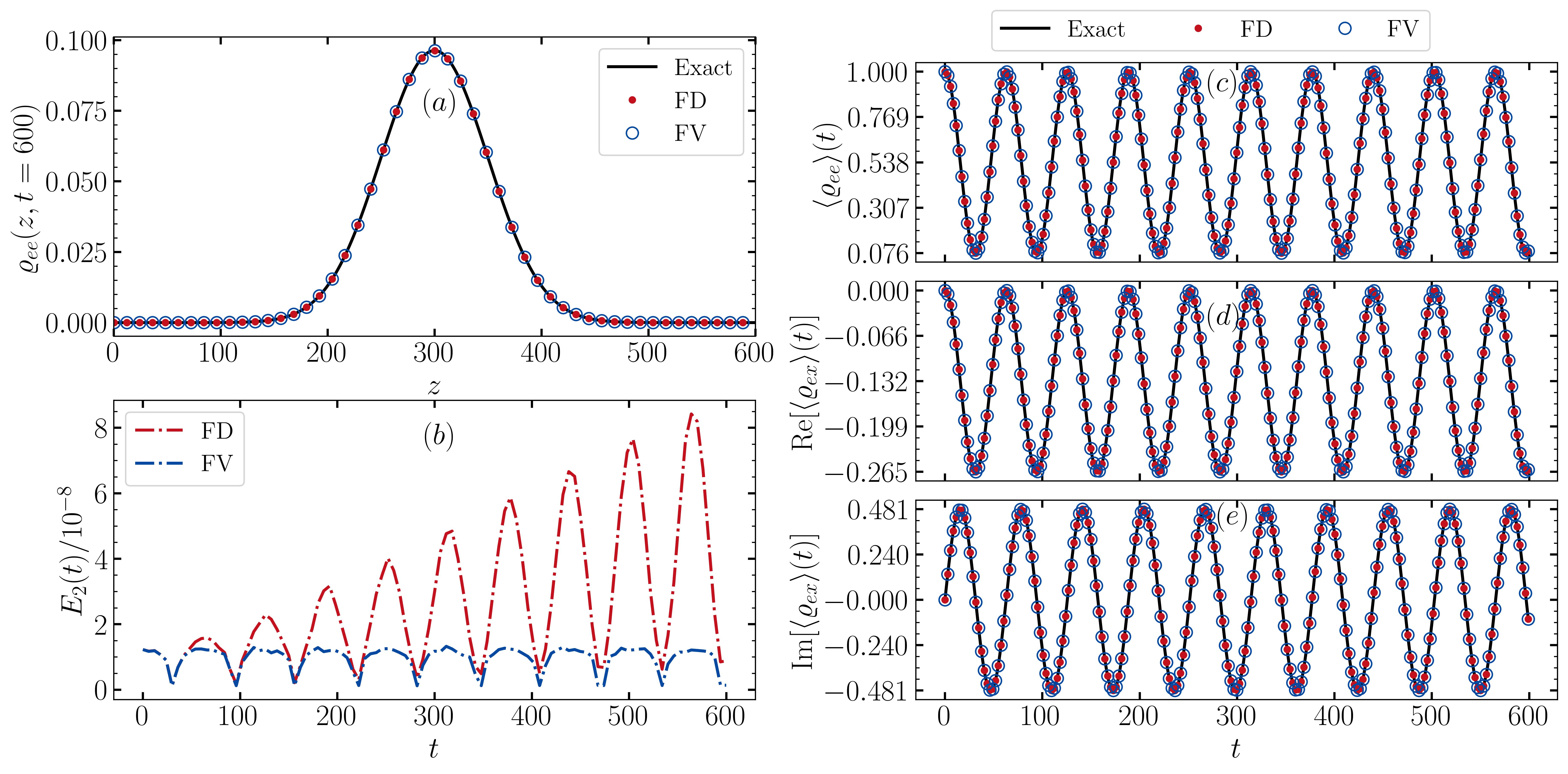}
    \caption{
    Results from the vacuum oscillation tests (see text for details). 
    Panel (a): comparison between the simulation and the exact solutions for $\R_\text{ee}$. 
    Panel (b): the error accumulation behavior for simulations with FD and FV schemes. 
    Panel (c)--(e): the time evolution of the quantities $\langle\R_\text{ee}\rangle,~ \text{Re}[\langle\R_{ex}\rangle]$ and  $\text{Im}[\langle\R_{ex}\rangle]$.
    }
    \label{fig:vacosc_test_eps}
\end{figure*}

Figure~\ref{fig:vacosc_test_eps} shows the results from vacuum oscillation tests. 
Panels (a) plots the spatial profiles of $\R_{ee}$ obtained from simulations with FD and FV on top of the analytical solution.  
Panel (b) shows the temporal evolution of the $E_2$ error associated with the same for both FD and FV. 
Similar to earlier results from the advection tests, 
$E_2$ remains at the level of $\mathcal{O}(10^{-8})$ for both FD and FV schemes.
Once again, the error grows faster in the FD case than FV. The envelope of the error for the latter remains almost constant, also due to the conservative nature of the FV scheme. The oscillatory behavior of the error accumulation shown in the figure is related to the oscillatory behavior of $\langle\R_{ee}\rangle(t)$: when $\langle\R_{ee}\rangle(t)$ approaches possible minimum value (0.076 for our choice of vacuum oscillation parameters), the profile $\R_{ee}(z, t)$ becomes smoother, resulting in relatively lower truncation error compared to the profile when $\langle\R_{ee}\rangle(t)=1$. 
Note that, here we have used the notation $\langle \R_{ab}\rangle(t)$ to indicate $\int \R_{ab}(z,t) dz/\int \R_{ee}^{t_0}(z) dz$.
For clarity, we show $\langle\R_{ee}\rangle(t)$, $\mathrm{Re}[\langle\R_{ex}\rangle(t)]$ and $\mathrm{Im}[\langle\R_{ex}\rangle(t)]$ respectively  in panels (c), (d) and (e). With this definition, what is shown in panel (c) corresponds to the time evolution of the survival probability of neutrinos. 
Once again, the results obtained here clearly match the exact solution well.

\section{Collective oscillations}
\label{sec:coll_osc_test}

In the previous two sections, we have performed tests for advection and vacuum oscillations. 
We now examine the results obtained from the simulations when collective oscillations occur. 
We switch off the vacuum oscillations and set $\mu=1$ so that $t$ and $z$ are expressed in terms of $\mu^{-1}$ hereafter.
The nonlinear and multi-angle nature of the $\nu\nu$ interaction makes the evolution of collective neutrino flavor conversions complicated. 
As a result, we cannot compare our simulation results to analytic solutions like in previous sections. 
Instead, we consider the deviation of several conserved quantities from their initial values to quantify the numerical errors. 

Following the convention used in Refs.~\cite{Joshua:2020,Wu-COSEnu-1:2021},  
we rewrite the density matrix $\R$ in the following way,
\begin{subequations}
\begin{equation}
    \R(t, z, v_z) \rightarrow g_\nu(v_z)\rho(t, z, v_z),
\end{equation}
and
\begin{equation}
    \bar\R(t, z, v_z) \rightarrow g_{\bar\nu}(v_z)\bar\rho(t, z, v_z),
\end{equation}
\end{subequations}

\noindent
where $g_{\nu(\bar\nu)}(v_z) = \frac{1}{4\pi^2 n_{\nu(\bar\nu)}}\int E^2dEf_{\nu(\bar\nu)}$ is the normalized neutrino (anti-neutrino) angular distribution function, i.e., $\int_{-1}^{1}g_{\nu(\bar\nu)}(v_z)dv_z = 1$. 
In this notation the angular distribution of electron lepton number (ELN) of the neutrinos takes the form $G(v_z) = g_\nu(v_z)-\alpha g_{\bar\nu}(v_z)$, where $\alpha=n_{\bar\nu_e}/n_{\nu_e}$. 
When the neutrino and antineutrino number densities are homogeneous in $z$, 
the length of the polarization vector $|\bm P|=1$ is conserved for any given $z$ and $v_z$ ($\bm P$ is defined by the decomposition $\rho = (P_0\mathcal{I} + \bm P\cdot\bm\sigma)/2$).
Furthermore, by virtue of the nature of the interaction, we can also show that the quantity 

\begin{equation}
    \bm{M_0}(t) = \int dz\int dv_z n_{\nu_e}(z, t)G(v_z)\bm{P}(t, z, v_z),
\end{equation}

\noindent
whose third component corresponds to the net neutrino lepton number, is also conserved when periodic boundary condition is considered~\cite{Bhattacharyya:2020}.

We have chosen the same angular distribution functions 

\begin{equation}
    g_{\nu(\bar\nu)}(v_z) \propto \text{exp}[-(v_z-1)^2/2\sigma_{\nu(\bar\nu)}^2],
\end{equation}

\noindent
as in Refs.~\cite{Joshua:2020,Wu-COSEnu-1:2021}. 
The corresponding neutrino ELN distributions for different values of $\alpha$ can be found in Ref.~\cite{Wu-COSEnu-1:2021}. 
For following discussion we consider $\alpha = 0.9$ as our fiducial value for the asymmetry parameter. 
We also use the values $\sigma_\nu = 0.6$ and $\sigma_{\bar\nu} = 0.5$ such that the $\bar\nu$ velocity distribution is more forward-peaked than that of $\nu$. 
Note that in the absence of the vacuum oscillation term, this configuration does not spontaneously produce any flavor transitions. 
Thus, one needs to provide initial perturbations to the off-diagonal components of the density matrix. 
In our tests, the perturbations are supplied in the following manner,

\begin{subequations}
\begin{equation}
    \rho_{ee}(z, v_z) = \left(1+\sqrt{1-\epsilon^2(z)}\right)/2 = \bar\rho_{ee}(z, v_z), 
    \label{eq:pert_ee}
\end{equation}
\begin{equation}
    \rho_{xx}(z, v_z)  = \left(1-\sqrt{1-\epsilon^2(z)}\right)/2 = \bar\rho_{xx}(z, v_z), 
    \label{eq:pert_xx}
\end{equation}
\begin{equation}
    \rho_{ex}(z, v_z)  = \epsilon(z)/2 = \bar\rho_{ex}(z, v_z),
    \label{eq:pert_ex}
\end{equation}
\label{eqs:perts}
\end{subequations}

\noindent
where $\epsilon(z) = 10^{-2}\text{exp}[-z^2/50]$.
We then impose the periodic boundary condition in $z$ and evolve the system with \cosenu. 

\begin{figure*}[ht]
    \includegraphics[width=\textwidth]{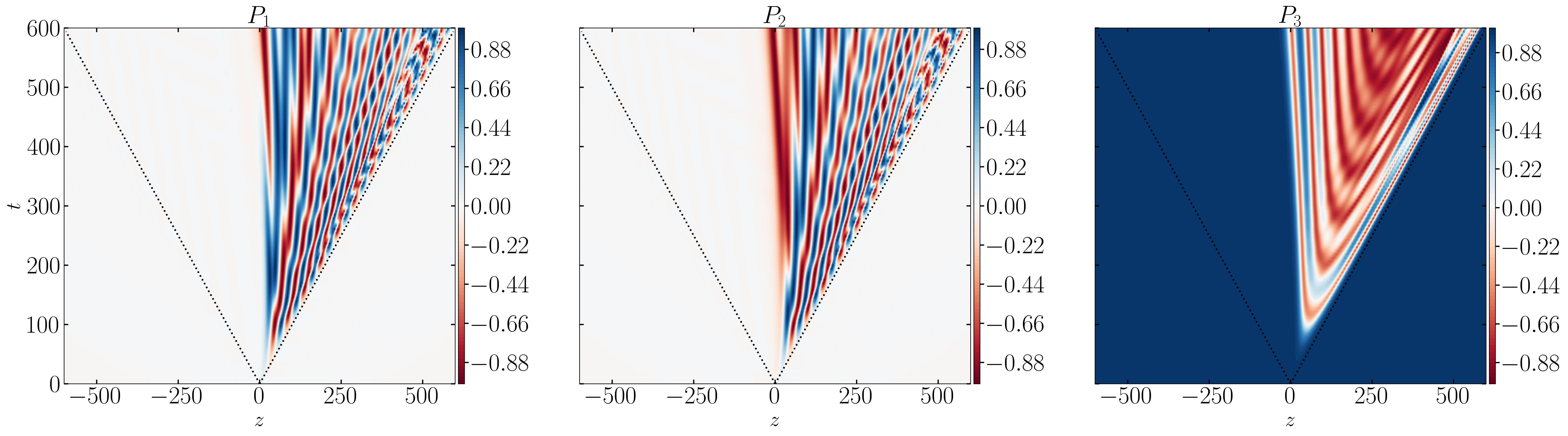}
    \caption{
    Evolution of the components of the polarization vector with $v_z=0.99$ for the fiducial case discussed in the sub-section \ref{sec:coll_osc_test} using $N_z = 2000$, $N_{v_z} = 100$ and $C_\text{CFL} = 0.2$. 
    The diagonal dotted lines are used to indicate the light cone.
    }
    \label{fig:collosc_Pi_cmap}
\end{figure*}

In Figure~\ref{fig:collosc_Pi_cmap}, we show the time evolution of the components of the polarization vector with $v_z=0.99$  for the above mentioned fiducial case. 
For this simulation we have chosen the parameters $N_z = 2000$, $N_{v_z} = 100$ and $C_\text{CFL} = 0.2$. 
We can see that collective neutrino oscillations, signified by the change of $P_i$ from their initial values, occur close to the center of the simulation domain where the initial perturbation is maximal. 
Flavor waves are produced and propagate primarily toward the positive $z$ direction, but never cross the light cone represented by the dotted lines.

\begin{figure*}[ht]
    \includegraphics[width=\textwidth]{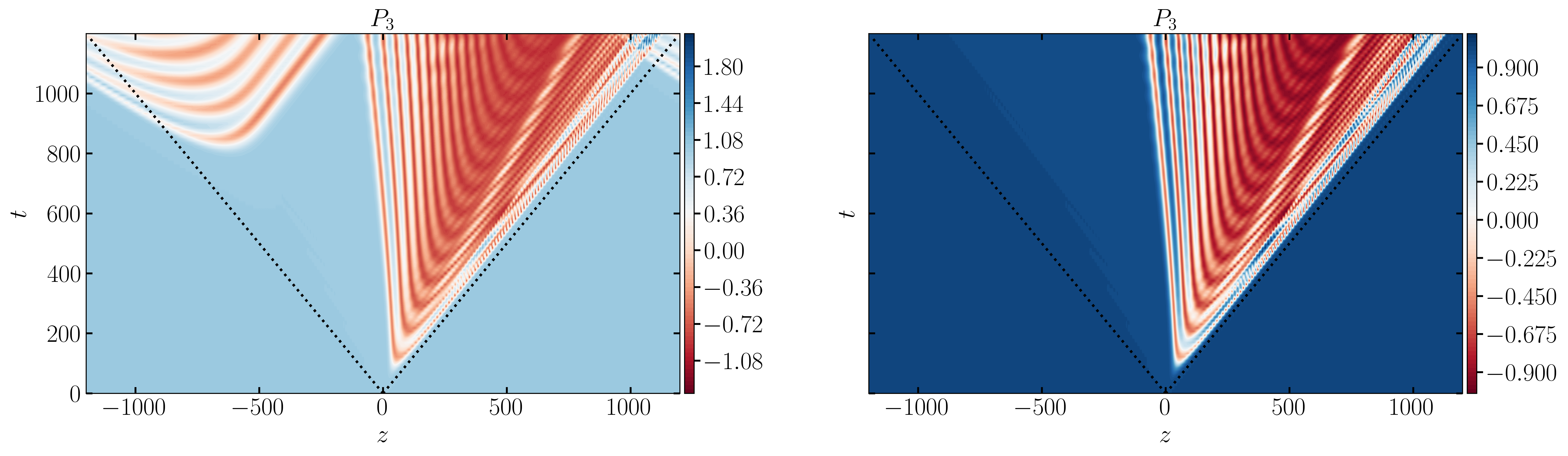}
    \caption{
    Comparison of FD results without (left panel) and with (right panel) the Kriess-Oliger dissipation scheme for the evolution of $P_3$. 
    The dotted lines are used to indicate the light cone.
    }
    \label{fig:collosc_noko_Pi_comparison_cmap}
\end{figure*}

Both FV and FD with Kriess-Oliger dissipation produces nearly identical results. 
In order to emphasize the importance of the KO dissipation scheme in FD, we additionally show in Figure~\ref{fig:collosc_noko_Pi_comparison_cmap} a comparison of $P_3$ with $v_z=0.99$  
obtained with and without Kriess-Oliger dissipation scheme. 
As can be seen from the left panel, the FD simulation without dissipation leads to numerical instabilities starting at $t\simeq 800$, indicated by the 
back-propagating pattern which violates the causality and produces unphysically large values of $P_3>1$. 

\begin{figure*}[h]
    \includegraphics[width=\textwidth]{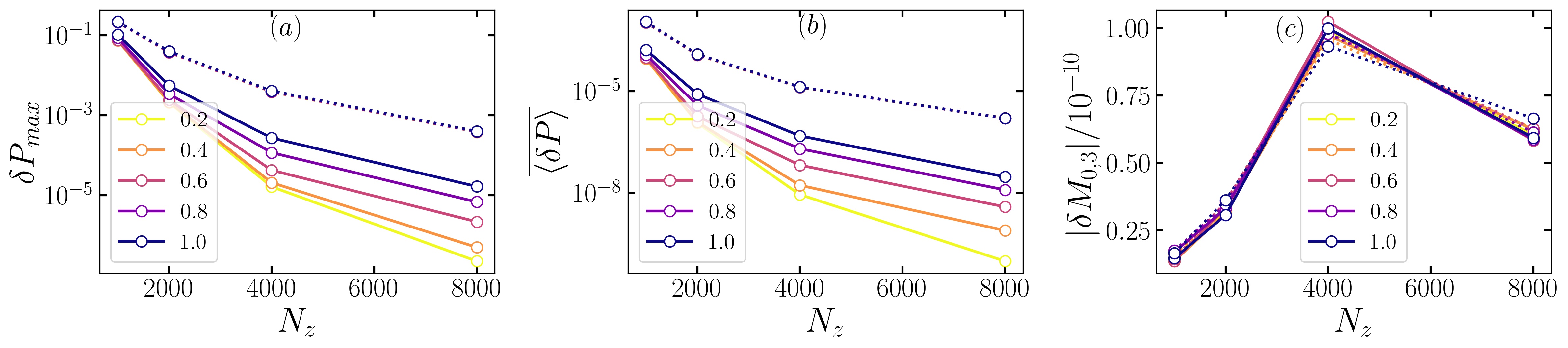}
    \caption{
    Error of the conserved quantities in \cosenu~ simulations for different values of $N_z$ and $C_\text{CFL}$. 
    Panel (a) shows the  $\delta P_\text{max}$, panel (b) shows the $\overline{\langle\delta P\rangle}$ and panel (c) shows the deviation of the third component of $\delta\bm M_0$. 
    Solid and dotted lines are for simulations with FV and FD schemes respectively.
    Different colors are used for results with different values of $C_\mathrm{CFL}$. Note here that the dotted lines for different values of $C_\mathrm{CFL}$ are indistinguishable in the sub-panels (a) and (b) as they are on top of each other.
    }
    \label{fig:collosc_dP_varying_Nz}
\end{figure*}

\begin{figure*}[h]
    \includegraphics[width=\textwidth]{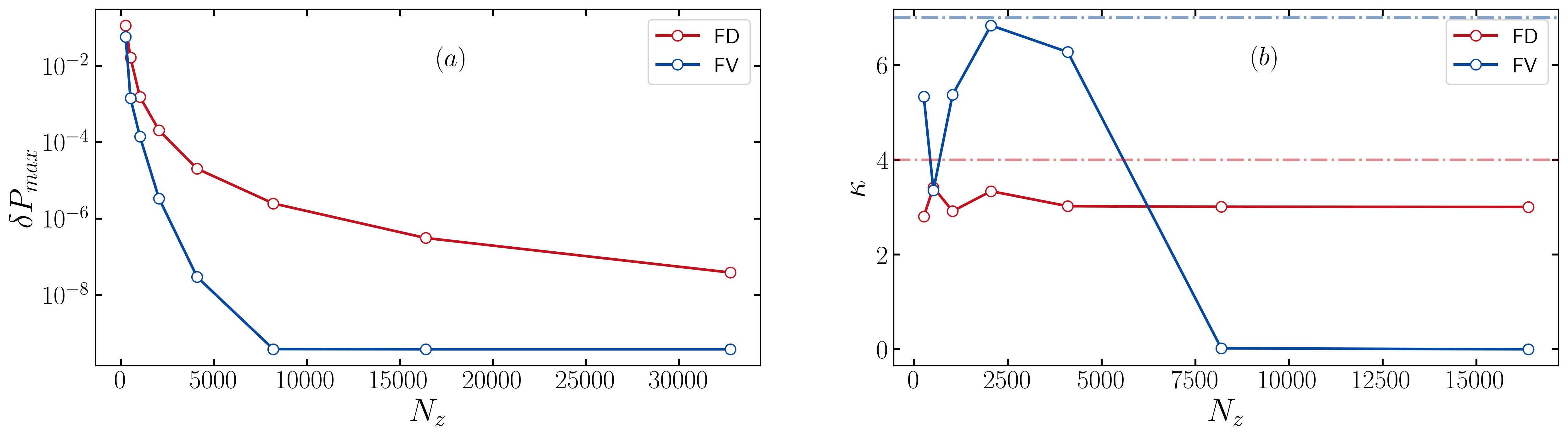}
    \caption{
    Values of $\delta P_{\rm max}$ [panel (a)] and $\kappa$ [panel (b)] obtained from \cosenu~ simulations for different values of $N_z$ while keeping $dt$ fixed. 
    The light dot-dashed blue and red lines in panel (b) represent the expected asymptotic values for the order of convergence.
    }
    \label{fig:kappa}
\end{figure*}

We now examine the numerical error accumulation in simluations for collective oscillations. 
We first check the maximum ($\delta P_\text{max}$) and the average ($\overline{\langle\delta P\rangle}$) of the deviation of the norm of the polarization vector from unity ($\delta P$). 
These quantities are defined as follows,

\begin{align}
    \delta P_{\max} =& \text {max} (|\delta P_{\nu}|_{\max}, ~|\delta P_{\bar\nu}|_{\max}), \\
    \overline{\langle\delta P\rangle} =& \frac{\int dz \int dv_z  |\delta P_\nu|g_\nu(v_z)}{\int dz \int dv_z g_\nu(v_z)},
\end{align}

\noindent
where $|\delta P_{\nu}|_{\max}$ and $|\delta P_{\bar\nu}|_{\max}$ are the maximum values of $|\delta P|$ among all velocity modes for $\nu$ and $\bar\nu$ respectively.
We then carry out the following two studies to  
understand the error accumulations from the spatial derivative and time integration separately. 
In the first one we vary the value of $C_\text{CFL}$ for each value of $N_z$ and in the second case we keep $\Delta t$ constant while varying $N_z$. 
The panels (a) and (b) of Figure~\ref{fig:collosc_dP_varying_Nz} show the results for the first case, where we plotted the errors obtained at the end of the tests. 
The panel (c) shows the deviation of the third component of $\bm M_0$ from its initial value. 
We have used dotted and solid lines to indicate the results from FD and FV respectively while different colors are used to indicate results obtained using different values of $C_\text{CFL}$.  
From panels (a) and (b), we can see that doubling the resolution leads to considerable improvements in the errors for both FD and FV. 
However, Changing the value of $C_\text{CFL}$ alone while keeping $N_z$ unchanged (equivalent to changing only $\Delta t$) does not affect the errors in the case of FD whereas the same improves the numerical accuracy of FV. 
This implies that in the case of FD, the major contribution to the error arises from the treatment of the spatial derivative while in the case of FV the error from the spatial part is similar to or slightly less than that from the time integration.
To further illustrate this we consider the second case in which we keep $\Delta t$ constant and choose a set of $N_{z}$ and $C_\text{CFL}$ as explained in the \ref{app:numerr}. $\Delta t$ here is chosen such that $N_z=32768$ has $C_\text{CFL}=1.0$. 
Figure~\ref{fig:kappa}(a) shows the values of $\delta P_\text{max}$ at the end of 1000 iterations from this test.
For FD, the error decreases monotonically as we increase the value of $N_z$. 
For FV, however, the error decreases until a saturation value $N_z^\text{sat}$($=8962$ here), after which it stays constant. 
The reason for this is that when $N_z$ reaches $N_z^\text{sat}$, the error from time integration become similar in magnitude as that from the spatial part. 
For $N_z$ larger than the saturation value, the predominant contribution to the error [see Eq.~(\ref{eq:err})] comes from the time integration, which is a constant since $\Delta t$ is a constant here.  
This is more evident from Figure~\ref{fig:kappa}(b) where we show the quantity $\kappa$, defined as 

\begin{equation}
\kappa = \frac{\text{log}\left[\delta P_\text{max}(N_{z, i}) /\delta P_\text{max}(N_{z,i+1})\right]} {\text{log}(2)},
\label{eq:kappa}
\end{equation}

\noindent
where $N_{z, i}$ refers to the $i$th $N_z$ value adopted in this set of calculations with $N_{z, {i+1}}=2N_{z, i}$. Since we kept $\Delta t$ constant, $\kappa$ corresponds to the order of numerical convergence when the error from time integration is negligible compared to the numerical error from the spatial derivative. If the situation is the other way around, $\kappa$ approaches zero. Thus, the approximate constant value of $\kappa$ for FD indicates the domination of the error from spatial derivative in the total error. 
Note here that $\kappa$ assumes similar value that we obtained for the order of accuracy in the Sec.~\ref{subsec:Advection}. On the other hand, $\kappa$ for FV initially increases with the increase in $N_z$ and then quickly approaches zero as a result of the aforementioned reasons.

\begin{figure*}[h]
    \includegraphics[width=\textwidth]{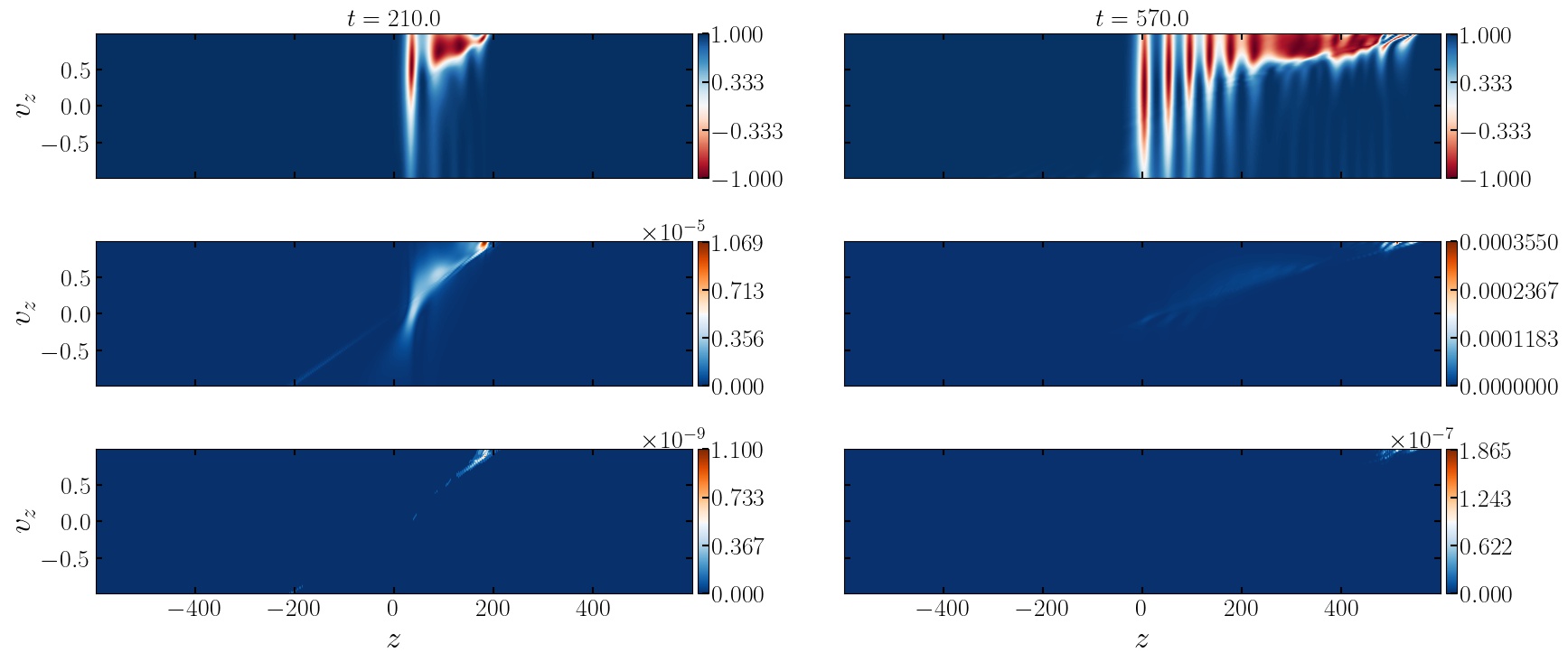}
    \caption{
    $P_3 (v_z,z)$ (top panels) at different times and the corresponding errors of $|\bm{P}|$
    with FD [middle] and FV [bottom] schemes.}
    \label{fig:err_dist}
\end{figure*}

From our simulations, we also observe that the $\delta P_\text{max}$ is associated with the collective mode which propagates with maximum speed.
This is because damping occurs most severely at the spatial regions where the smoothness is the lowest (see also the previous advection tests in Sec.~\ref{subsec:Advection}). 
This is illustrated in Figure~\ref{fig:err_dist} where the error distribution as a function of $z$ and $v_z$ is plotted. 
The top panels show $P_3(z,v_z)$ at two different times and serve as a reference to locate the collective wave fronts. 
The middle and the bottom panels show the deviation $\delta P(z,v_z)$ obtained from FD and FV, respectively. 
This plot clearly shows that the errors are larger at the edge of the wave fronts with largest values occurring at $v_z\simeq 1$ as discussed above.

\begin{figure*}[h]
    \includegraphics[width=1\textwidth]{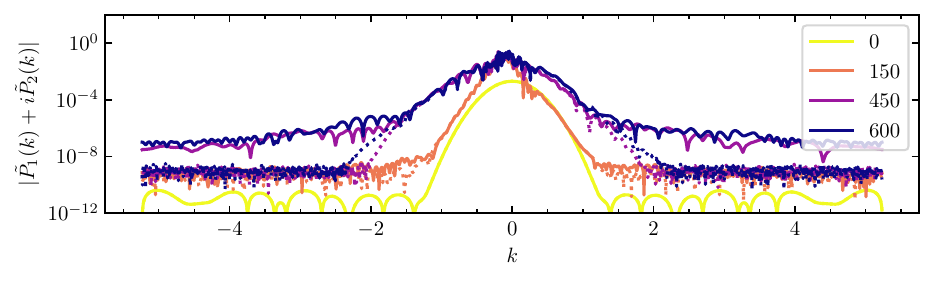}
    \caption{
    The Fourier spectrum computed using the perpendicular components of the polarization vector at different times indicated by different colors
    obtained with the FD (dotted lines) and FV (solid lines) schemes. 
    }
    \label{fig:coll_power_spectrum_0.9}
\end{figure*}

Finally, in Figure~\ref{fig:coll_power_spectrum_0.9} we show the time evolution of the Fourier spectra for $v_z=0.99$ obtained from our fiducial simulation ($N_z=2000$ and $C_\text{CFL}=0.2$) with FD (dotted lines) and FV (solid lines). 
Both FD and FV produce similar Fourier spectra in the range of $-1 \leq k\leq 1$. 
For larger values of $|k|$, we see that the FV scheme results in faster rise in spectrum when compared to the case with FD. 
Nevertheless, we confirm that this difference does not lead to any discernible impact on other quantities examined here or in Ref.~\cite{Wu-COSEnu-1:2021}. 
We note that the reason for causing this difference likely differs from that discussed in Sec.~\ref{subsec:Advection} for advection tests, but we do not pursue the exact reason in this paper and leave it for future work.

\section{Performance comparison}
\label{sec:performance}
In the previous sections we have discussed the implementation details and the numerical behavior of \cosenu. 
We now briefly comment on the parallelization strategy as well as the computational performance of the code. \cosenu~exploits multi-core and GPU accelerations with the help of the directive based parallelization provided by OpenMP~\cite{OpenMP} and OpenACC~\cite{OpenACC}, respectively, with a single source code. For OpenMP, we extensively used \texttt{parallel for collapse(n)} clause for every \texttt{n}-level tightly nested loop without data dependency, which is most of our case. The \texttt{reduction} clause is appended to the parallel directives whenever necessary (mostly while computing integrals and doing the analysis). Similarly in OpenACC, we extensively used \texttt{parallel loop collapse(n)} clause for tightly nested loops and \texttt{reduction} clause in analysis routines. In the computation of the the right-hand side if the Eq.~\eqref{eq:eom}, we further exploit the levels of parallelism such as \texttt{gang}, \texttt{worker} and \texttt{vector} provided by OpenACC. For data management on GPU, we currently rely on the feature of unified memory allowing a compiler handle the data allocation on the GPU and its movement between the CPU when necessary. Using the minimal set of the compiler directives mentioned above already attain acceptable time-to-solution for the scale of problems in this work and serve as the baseline for future extension. We note that the performance can be optimized by more sophisticated controls on data movement and asynchronous execution, which will be explored in future iterations of \cosenu.

\begin{figure*}[ht]
    \centering
    \includegraphics[width=\textwidth]{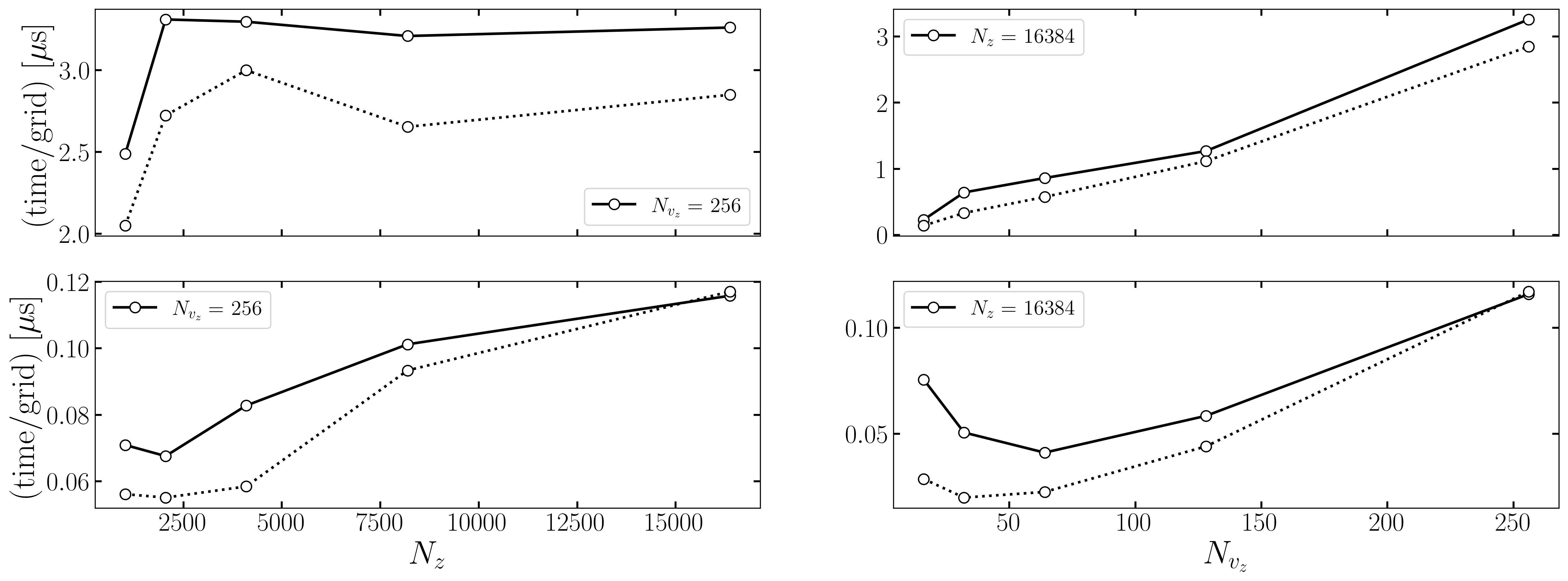}
    \caption{Comparison of the average computational time taken by \cosenu~package per grid per iteration for different  grid configurations. Left panels display the time per grid when $N_{v_z}$ is kept constant and the right panels shows the same when $N_z$ is kept constant. 
    The top and bottom panels show the multi-core and GPU performances respectively. Note that Solid and dotted lines are used to differentiate the results with FV and FD respectively and different colors for the lines are used to indicate different values of $N_{v_z}$.}
    \label{fig:walltimes}
\end{figure*}

Figure~\ref{fig:walltimes} compares the computational time per iteration per grid number ($N_z\times N_{v_z}$) for both FD and FV using different values of $N_z$ and $N_{v_z}$. 
The left and right panels show the results obtained with fixed $N_{v_z}$ and $N_z$, while the top and the bottom panels are with a 20 core Intel\textregistered Xeon\textregistered E5-2620 processor and NVIDIA P100 GPU, respectively. Solid and dotted lines are used to distinguish the results from FV and FD. Note that here we have switched off all the parts which analyze and output the simulation results in the code, but only considered the computational part, which consumes the major part of the total computing time. First, for nearly all cases the FD version performs slightly better than the FV version on both the GPU and CPU (note that the FV version provides better accuracy as described in previous sections).
Second, it is clear that the performance with a single GPU card used here provides much better ($\gtrsim 20$ times faster) than that with a single CPU node. 
Third, the cases with constant $N_{v_z}$ show nearly perfect scaling when varying $N_z$ in CPU. The scaling is slightly less than ideal in GPU mainly due to tests performed in the under-saturated region of $N_z$. On the other hand, the computing time per grid per iteration increases nearly linearly as $N_{v_z}$ becomes large, as shown in the right panels of Figure~\ref{fig:walltimes}. The linear dependency is consistent with the estimated computing costs of the integration over $N_{v_z}$ in the right-hand side of Eq.~\eqref{eq:eom}, which grows linearly over $N_{v_z}$ and is the most time-consuming part of the program.


\section{Conclusion and future plans}
\label{sec:discussion}
We have provided details of the numerical implementation in the simulation code, the Collective Simulation Engine for Neutrinos (\cosenu), which numerically solves a set of partial differential equations that dictates the dynamics of the collective neutrino flavor conversions in 1+1+1 dimensions. In-depth details for both the finite difference method supported by the third order Kriess-Oliger dissipation scheme as well as the finite volume method with seventh order weighted essentially non-oscillatory scheme were discussed.
We have also tested the code against advection and vacuum oscillations and shown that \cosenu~is capable of reproducing the analytical results to a very good precision.

For collective neutrino oscillations, 
we discussed a fiducial case and demonstrated that adopting the dissipation scheme in the finite difference version of the  implementation is essential to prevent the growth of numerical instabilities.
The analysis of the conserved quantities showed that \cosenu~can simulate collective oscillations with very small numerical errors when appropriate spatial and temporal resolutions are chosen.
We have also evaluated and provided the performance of \cosenu~on both CPUs and GPU.
The public version of the \cosenu~package is available at \url{https://github.com/COSEnu/COSEnu}.

Beyond what was described in this work, we plan to extend \cosenu~to include other spatial and phase-space dimensions,  as well as the collisions between neutrinos and matter.
We will also explore other numerical schemes to further suppress the associated errors, and pursue better speed-up with cross nodes and/or multiple GPU cards. 
All these improvements will be released in future version of \cosenu.

\section*{Acknowledgment}
\label{sec:acknowledgment}
M.-R.~W. and M.~G. acknowledge supports from the Ministry of Science and Technology, Taiwan under Grant No.~110-2112-M-001-050, No.~111-2628-M-001-003-MY4, and the Academia Sinica under Project No.~AS-CDA-109-M11.
M.-R.~W. also acknowledges supports from the Physics Division, National Center for Theoretical Sciences, Taiwan.
M.~G. and M.-R.~W. appreciate the computing resources provided by the Academia Sinica Grid-computing Center.
C.-Y.~L. thanks the National Center for High-performance Computing (NCHC) for providing computational and storage resources.
Z.~X. acknowledge supports from the European Research Council (ERC) under the European Union's Horizon 2020 research and innovation programme (ERC Advanced Grant KILONOVA No.~885281). Authors would also like to thank Taiwan Computing Cloud and NVIDIA for organizing GPU hackathon and also would like to acknowledge the help from Mathew Colgrove and Mason Wu of NVIDIA.


\appendix
\section{Seventh order WENO-Explicit form}
\label{app:weno7_explicit}

\begin{figure*}[h]
    \centering
    \includegraphics[width=0.9\textwidth]{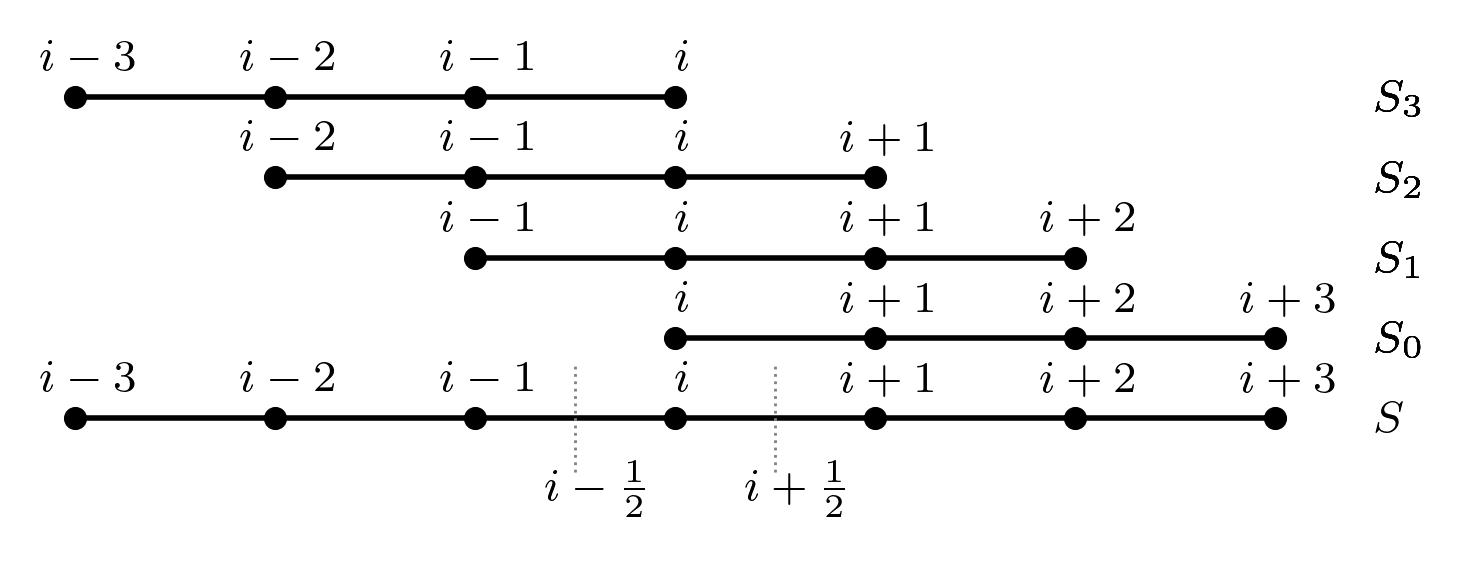}
    \caption{The pictorial representation of the stencils used to compute the fluxes at the locations $x_i-\Delta x/2$ and $x_i+\Delta x/2$ (locations of these points on the stencil is represented using vertical dotted lines). The stencil labelled $S$ corresponds to the base stencil. The WENO scheme implemented in the \cosenu~ uses the linear combination of 4th order accurate fluxes computed using sub-stencils labelled $S_r,~ r=0, 1, 2, 3$.}
    \label{fig:stencils}
\end{figure*}

In order to reconstruct the flux at $i+1/2$, we consider a stencil $S = \{i-3, i-2, i-1, i, i+1, i+2, i+3\}$ and divide it  into four sub-stencils of consists of four grid points each (see the Fig. \ref{fig:stencils} for a pictorial representation of the stencil $S$ and the sub-stencils used for the WENO reconstruction)\cite{Shu:1998}.

\begin{eqnarray*} 
        S_0 &=& \{i, i+1, i+2, i+3\}, ~~\textrm{for~} r=0, \\ 
        S_1 &=& \{i-1, i, i+1, i+2\}, ~~\textrm{for~} r=1, \\
        S_2 &=& \{i-2, i-1, i, i+1\}, ~~\textrm{for~} r=2, \\
        S_3 &=& \{i-3, i-2, i-1, i\}, ~~\textrm{for~} r=3. 
\end{eqnarray*}
\noindent
Then we have 

\begin{equation}
\begin{split}
    \label{eq:f7}
    \tilde f^{(7)}_{i+1/2} = & (\frac{-1}{140})\bar f_{i-3} + (\frac{5}{84})\bar f_{i-2} + (\frac{-101}{420})\bar f_{i-1} + \\& (\frac{319}{420})\bar f_{i} + (\frac{107}{210})\bar f_{i+1} + (\frac{-19}{210})\bar f_{i+2} + (\frac{1}{105}) \bar f_{i+3},
\end{split}
\end{equation}

\noindent
and 

\begin{subequations}
    \begin{equation}
        \hat f^{(4)}_{r = 0, i+1/2} = (\frac{3}{12})\bar f_{i} + (\frac{13}{12})\bar f_{i+1} + (\frac{-5}{12})\bar f_{i+2} + (\frac{1}{12})\bar f_{i+3}, 
    \end{equation}
    \begin{equation}
        \hat f^{(4)}_{r = 1, i+1/2} = (\frac{-1}{12})\bar f_{i-1} + (\frac{7}{12})\bar f_{i} + (\frac{7}{12})\bar f_{i+1} + (\frac{-1}{12})\bar f_{i+2},
    \end{equation}
    \begin{equation}
        \hat f^{(4)}_{r = 2, i+1/2} = (\frac{1}{12})\bar f_{i-2} + (\frac{-5}{12})\bar f_{i-1} + (\frac{13}{12})\bar f_{i} + (\frac{3}{12})\bar f_{i+1}, 
    \end{equation}
    \begin{equation}
        \hat f^{(4)}_{r = 3, i+1/2} = (\frac{-3}{12})\bar f_{i-3} + (\frac{13}{12})\bar f_{i-2} + (\frac{-23}{12})\bar f_{i-1} + (\frac{25}{12})\bar f_{i}.
    \end{equation}
\label{eq:f4s}
\end{subequations}

 From Eqs~.(\ref{eq:fn_lin_comb}, \ref{eq:f7}, \ref{eq:f4s}), we get $d_0 = \frac{1}{35},~d_1 = \frac{12}{35},~d_2 = \frac{18}{35}$ and $d_3 = \frac{4}{35}$. Using these values of $d_r$ we can construct the weight factors using Eq.~(\ref{eq:weight}). Furthermore, for each sub-stencil, one need to estimate the smoothness indices (SI$_r$). In our implementation we have used the following SI$_r$ \cite{Balsara:2000, Jiang:1996} for $r=0, 1, 2, 3$.
 
 \begin{subequations}\label{eq:Si4}
    \begin{equation}
         \begin{split}
             \textrm{SI}_0 &= \bar f_i(2107\bar f_i - 9402\bar f_{i+1} + 7042\bar f_{i+2} - 1854\bar f_{i+3}) 
             \\& + \bar f_{i+1}(11003\bar f_{i+1} - 17246\bar f_{i+2} + 4642\bar f_{i+3}) 
             \\& + \bar f_{i+2}(7043\bar f_{i+2} - 3882\bar f_{i+3}) + 547\bar f_{i+3}^2,
         \end{split}
     \end{equation}
     \begin{equation}
        \begin{split}
            \textrm{SI}_1 &= \bar f_{i-1}(547 \bar f_{i-1} - 2522 \bar f_i + 1922 \bar f_{i+1} - 494 \bar f_{i+2}) 
            \\& + \bar f_i (3443 \bar f_i - 5966 \bar f_{i+1} + 1602 \bar f_{i+2}) \\& + \bar f_{i+1} (2843 \bar f_{i+1} - 1642 \bar f_{i+2}) + 267 \bar f_{i+2}^2,
        \end{split}
    \end{equation}
    \begin{equation}
        \begin{split}
            \textrm{SI}_2 &= \bar f_{i-2}(267 \bar f_{i-2} - 1642 \bar f_{i-1} + 1602 \bar f_{i} - 494 \bar f_{i+1}) 
            \\& + \bar f_{i-1} (2843 \bar f_{i-1} - 5966 \bar f_{i} + 1922 \bar f_{i+1}) \\
            &+ \bar f_{i} (3443 \bar f_{i} - 2522 \bar f_{i+1})+ 547 \bar f_{i+1}^2,
        \end{split}
    \end{equation}
\begin{equation}
    \begin{split}
        \textrm{SI}_3 &= \bar f_{i-3} (547 \bar f_{i-3} - 3882 \bar f_{i-2} + 4642 \bar f_{i-1} - 1854 \bar f_{i}) 
        \\& + \bar f_{i-2} (7043 \bar f_{i-2} - 17246 \bar f_{i-1} + 7042 \bar f_{i}) \\
        &+ \bar f_{i-1} (11003 \bar f_{i-1} - 9402 \bar f_{i}) + 2107 \bar f_i^2.
    \end{split}
    \end{equation}
 \end{subequations}

Simple pseudo code for the above described steps for calculating the value of the flux at $x_i + \Delta x/2$ is given in Algorithm~\ref{algo:weno7}.

\begin{algorithm}[H]
\caption{Estimation of flux at $i+1/2$ using 7th order accurate WENO scheme}
\begin{algorithmic} [1]
\label{algo:weno7}
\STATE For each spatial grid $i$ assign:\\
        \tab $S_0 \gets \{\bar f_i, \bar f_{i+1}, \bar f_{i+2}, \bar f_{i+3}\}$ \\
       \tab $S_1 \gets \{\bar f_{i-1}, \bar f_{i}, \bar f_{i+1}, \bar f_{i+2}\}$ \\
       \tab $S_2 \gets \{\bar f_{i-2}, \bar f_{i-1}, \bar f_{i}, \bar f_{i+1}\}$ \\
       \tab $S_3 \gets \{\bar f_{i-3}, \bar f_{i-2}, \bar f_{i-1}, \bar f_{i}\}$ 
\STATE Compute 4th order accurate values $\hat f^{(4)}_{r, ~i + 1/2}$ for each $S_r, r=0, 1, 2, 3$ using Eq.~\eqref{eq:f4s}
\STATE Compute smoothness \textrm{SI}$_r$ for each $S_r$ using Eq.~\eqref{eq:Si4}
\STATE Compute the weight factor $w_r$ using $d_r$ and \textrm{SI}$_r$ using Eq.~\eqref{eq:weight}
\STATE Compute average weight factor $\bar w_r = \frac{w_r}{\Sigma_r w_r}, r=0, 1, 2,3$
\STATE Compute the 7th order accurate values at $\hat f^{(7)}_{i + 1/2}$ using $\bar w_r$ and $\hat f^{(4)}_{r, ~i+1/2}$ with Eq.~\eqref{eq:fn_w_lin_comb}
\end{algorithmic}
\end{algorithm}

 We can follow the similar procedure to reconstruct the value of flux at $i-1/2$ as well.
 
 \section{Effect of Kreiss-Oliger dissipation}
\label{app:KO}
\begin{figure*}[h]
    \centering
    \includegraphics[width=\textwidth]{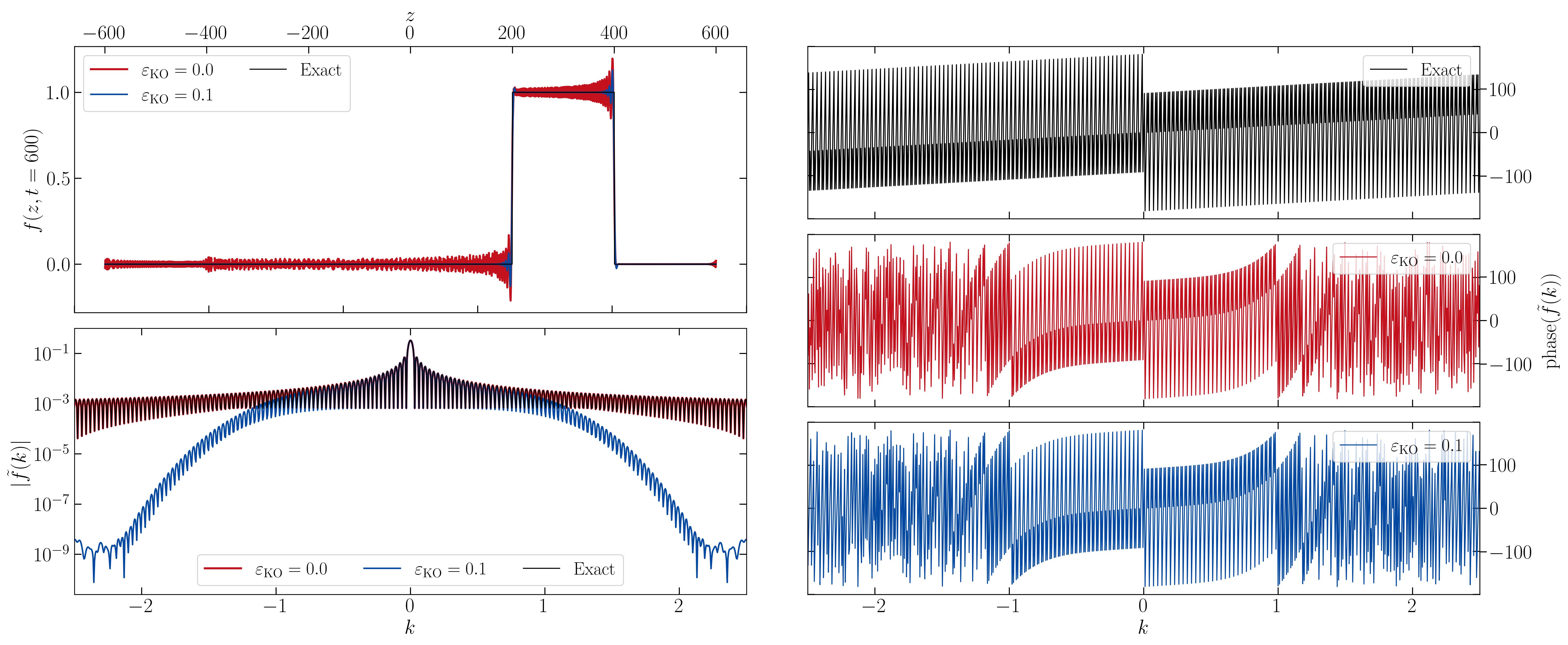}
    \caption{Comparison of the effect of KO dissipation scheme. Top-left panel shows results from the pure advection for a box profile with (blue) and without (red) Kreiss-Oliger dissipation. The solid black line represents the exact solution. Bottom-left panel displays the absolute value of the discrete Fourier spectra corresponding to the profiles shown in the top left panel and the right panels show the phase values (in degrees) of the same Fourier spectra.}
    \label{fig:adv_w_and_wo_ko}
\end{figure*}

As discussed in Sec.~\ref{sec:numerics}, our implementation of the \cosenu~simulation with the FD method uses the third order KO dissipation scheme to take care of the possible numerical instabilities. 
To illustrate the advantage of the KO scheme, in Figure~\ref{fig:adv_w_and_wo_ko} we show the results and the Fourier transform of the results from the FD simulations with and without KO dissipation. 

As shown in the bottom left panel of Figure~\ref{fig:adv_w_and_wo_ko}, the Fourier power spectrum from the simulation without KO dissipation has the same amplitude as that of the exact solution. 
However, the phases are only retained relatively well for $|k|\lesssim 1$ but become random for $|k|\gtrsim 1$.
This leads to a numerical instability clearly visible in the top left panel.
When we apply the KO scheme with $\varepsilon_{\rm KO}=0.1$, although the phases remain random for $|k|\gtrsim 1$, it clearly suppresses the 
amplitudes of these modes and thus the corresponding numerical artifact.

\section{Numerical estimation of the order of accuracy}
\label{app:numerr}
The general form of the numerical truncation error, $E$, can be expressed as,
\begin{equation}
    E\simeq C_z \Delta z^\alpha + C_t \Delta t^\beta,
\end{equation}

\noindent
where $C_z$ and $C_t$ are some constants and $\alpha$ and $\beta$ are the spatial and temporal order of accuracy respectively. 
Now, if we have a set $S=\left\{\Delta z_0, \Delta z_1,... | \Delta z_{i+1}=\frac{\Delta z_i}{r}, r>1\right\}$, then

\begin{equation}
    E_i = C_z(\frac{\Delta z_0}{r^i})^\alpha + C_t \Delta t_i^\beta.
    \label{eq:err}
\end{equation}
If we choose a set of $C_\text{CFL}$ such that $\Delta t_i$ is a constant for all values of $\Delta z_i\in S$ then the spatial order of accuracy $\alpha$ can be estimated using

\begin{equation}
    \alpha = \frac{\text{log}\big((E_{i+1}-E_i)/(E_{i+2}-E_{i+1})\big)}{\text{log}(r)}.   
\end{equation}
\bibliographystyle{elsarticle-num}
\bibliography{main_2.bib}
\end{document}